\documentclass{llncs}
\usepackage [T1]{fontenc}

\usepackage[cmtip,all]{xy}
\usepackage{mathrsfs,amssymb}
\usepackage{mathtools}
\usepackage[ruled,vlined]{algorithm2e}

\newcommand{\F}{\mathbb{F}}

\begin{document}
\title{Cover attacks for elliptic curves with prime order}
\author{Song Tian\thanks{Song Tian is supported by the National Natural Science Foundation of China under Grant No. 61802401.}\\  tiansong@iie.ac.cn}
\institute{State Key Laboratory of Information Security, Institute of Information Engineering, Chinese Academy of Sciences, Beijing, 100093, China
}
\date{}
\maketitle
\begin{abstract}
We give a new approach to the elliptic curve discrete logarithm problem over cubic extension fields $\mathbb{F}_{q^3}$. It is based on a transfer: First an $\mathbb{F}_q$-rational $(\ell,\ell,\ell)$-isogeny from the Weil restriction of the elliptic curve under consideration with respect to $\mathbb{F}_{q^3}/\mathbb{F}_q$ to the Jacobian variety of a genus three curve over $\mathbb{F}_q$ is applied and then the problem is solved in the Jacobian via the index-calculus attacks. Although using no covering maps in the construction of the desired homomorphism, this method is, in a sense, a kind of cover attack. As a result, it is possible to solve the discrete logarithm problem in some elliptic curve groups of prime order over $\mathbb{F}_{q^3}$ in a time of $\tilde{O}(q)$.
\keywords{cover attacks, discrete logarithm problem, Weil restriction, isogeny, Jacobian variety }
\end{abstract}

\section{Introduction}
It is well known that the discrete logarithm problem in the groups of rational points of elliptic curves over finite fields can be solved in a time which is given by the square root of the group order. The corresponding algorithms in fact work in all finite groups and are therefore called generic algorithms. Up to now the fastest algorithms for the discrete logarithm problem in most of elliptic curves are indeed the generic algorithms. Additionally, it is easy to construct curves which are suited for cryptographic applications. For this reason, elliptic curve cryptography is now widely used in practice, in particular for high-security applications.

On the other hand, there are now some classes of elliptic curves known for which the elliptic curve discrete logarithm problem can be solved faster than generic algorithms. In particular, it is well known that some elliptic curves over finite non-prime fields can be attacked with what is called \emph{cover attack} by C.~Diem (see for example \cite{cover-attack-report}).

Let $E$ be an elliptic curve over an extension field $\mathbb{F}_{q^n}$. The idea of cover attacks is to reduce the discrete logarithm problem to the corresponding problem in the Jacobian of a curve $C$ of genus $g\ge n$ over $\mathbb{F}_q$, where it can be solved in a time of $\tilde{O}(q^{2-2/g})$ via the index calculus method \cite{doubleLP}. In order that the attack is faster than generic attacks, one wants that the genus of $C$ is small. For example, in the case that  $E(\mathbb{F}_{q^n})$ is prime (which is often the case in cryptographic applications), it is optimal if $g$ is equal to the extension degree $n$.

The first application of this idea is the \emph{GHS attack} \cite{diemGHS,GHS,Hess-GHS-rev}. This attack is only of relevance for relatively few curves. For example, if $q$ is odd and $n$ is prime and not 3, 5 or 7, the genus of the resulting curve is always so large that one can say that the attack fails. For $n=3$, if we require $C$ to be of genus $3$, then a necessary condition is that the elliptic curve has three rational points of order $2$ \cite{diemGHS,Momose}. This therefore raises the natural question whether elliptic curves over $\mathbb{F}_{q^3}$ with cofactor $1$ or $2$ can be attacked with cover attacks. A recent discussion in \cite{tian2018} shows that $C$ needs to be non-hyperelliptic if the elliptic curve has prime order, and that when the cofactor is $2$, it might be possible to construct a genus $3$ hyperelliptic curve $C$ together with a degree $3$ morphism $C\to E$.

In this work, we give a new approach to the problem. The main idea is to attempt to compute $(\ell,\ell,\ell)$-isogenies from the Weil restriction of an elliptic curve to the Jacobian of a genus $3$ curve. This is motivated by the fact that principally polarized abelian varieties of dimension $3$ are the generalized Jacobian varieties \cite{Oort}. If we take such an isogeny for an elliptic curve $E$ of prime order, we can transfer the discrete logarithm problem from $E(\mathbb{F}_{q^3})$ to the Jacobian of a non-hyperelliptic curve over $\F_q$, and solve it there in a time of $\tilde{O}(q)$ \cite{diem_nonhyperelliptic}. In contrast, if we apply the index-calculus method based on the summation polynomials to the discrete logarithm problem in $E(\mathbb{F}_{q^3})$, then the complexity is
$\tilde{O}(q^{4/3})$ \cite{summation}.

We note that the GHS attack \cite{diemGHS,GHS} and the work \cite{tian2018} transfer the discrete logarithm problem in a way which relies on
the covering map: The homomorphism from the group under consideration to another group is the composition of
the pullback homomorphism induced by the covering map and the trace map (or at the level of function fields of curves the composition of the conorm and norm maps). This contrasts with the direct construction of a
homomorphism in the present work.

The method of computing the $(\ell,\ell,\ell)$-isogenies is basically the same as the one in \cite{lllisogeny}. The only difference is that the isogenies start from the Weil restriction of an elliptic curve instead of the Jacobian of a hyperelliptic curve. We remark here that the most important notion for the computation is normal Weil set, which is introduced by Shepherd-Barron in \cite{Barron}. First, we use Weil sets on the Weil restriction to construct a normal Weil set on
the quotient of the Weil restriction by a rational subgroup with certain property, from which we can compute an equation for the resulting curve. The question then is, how to compute the image in the Jacobian of this curve of a given point on the Weil restriction?
For this, we build a link between the normal Weil set on the quotient and a normal Weil set on the Jacobian. Doing so allows us to know the values of functions in the latter normal Weil set at the image point. Then we recover the image point from these values by solving a system of polynomial equations.

The paper is organized as follows. In section 2, we give an introduction to normal Weil set. In section 3, we give a description of the algorithm for computing $(\ell,\ell,\ell)$-isogenies. In section 4, we give a conclusion.


\section{Normal Weil sets}
The notion of normal Weil set is introduced by Shepherd-Barron in \cite{Barron} as the algebraic analogue of the ratios of theta functions of level $N$. In this section we recall normal Weil set and explain how to construct Weil sets on the Weil restriction of an elliptic curve and on the Jacobian variety of a curve.

Let $\mathcal{A}$ be an abelian variety over an algebraically closed field $k$ and $W$ a symmetric divisor on $\mathcal{A}$. Let $N$ be a positive integer. Then for each $N$-torsion point $P\in \mathcal{A}(k)$, there
is a rational function $f_P$ whose divisor is $NT_P^*W-NW$, where $T_P:\mathcal{A}\to \mathcal{A}$ is the translation by $P$. We call $f_P$ a \emph{Weil function} for $P$, and
call an ordered set $\{f_P\}_{P\in \mathcal{A}[N]}$ of Weil functions a \emph{Weil set}. The set of all pairs $(P,f_P)$ with the operation given by $(P,f_P)(Q,f_Q)=(P+Q,f_PT_P^*f_Q)$ is
a group. The Weil pairing $e_N:\mathcal{A}[N]\times \mathcal{A}[N]\to \mu_N$ can be rephrased as taking the commutator; more precisely,
$e_N(P,Q)=\frac{f_PT_P^*f_Q}{f_QT_Q^*f_P}$.

Fix a bilinear pairing $d_N$ on $\mathcal{A}[N]$ such that $e_N(P,Q)=d_N(P,Q)/d_N(Q,P)$ for all $P,Q\in \mathcal{A}[N]$. We say that a Weil set $\{f_P\}_{P\in \mathcal{A}[N]}$ is \emph{normal} if for
all $P,Q\in \mathcal{A}[N]$, we have $f_PT_P^*f_Q=d_N(P,Q)f_{P+Q}$. Given a Weil set $\{f_P\}_{P\in \mathcal{A}[N]}$, we can use Algorithm \ref{algorithm_for_normalization} to compute an ordered set $\{\alpha_P\}_{P\in \mathcal{A}[N]}$ of scalars
such that $\{\alpha_Pf_P\}_{P\in \mathcal{A}[N]}$ is normal (see \cite[Section 3.2]{Barron} for details).

\begin{algorithm}\label{algorithm_for_normalization}
 \KwData{a Weil set $\{f_P\}_{P\in \mathcal{A}[N]}$ ($N$ is a prime, $f_0=1$), a bilinear pairing $d_N$ on $\mathcal{A}[N]$ such that $e_N(P,Q)=d_N(P,Q)/d_N(Q,P)$ }
 \KwResult{ an ordered set $\{\alpha_P\}_{P\in \mathcal{A}[N]}$ of scalars such that $\{\alpha_Pf_P\}_{P\in \mathcal{A}[N]}$ is normal. }

 \nl Compute a basis $P_1,\ldots,P_{2g}$ of $\mathcal{A}[N]$\; \tcc{g is the dimension of $\mathcal{A}$}
 \nl $\alpha_0=1$\;
 \nl \For{$i=1$ \KwTo $2g$}{compute $u_i=\prod_{j=1}^{N-1}d_N(P_i,jP_i)\frac{f_{(j+1)P_i}}{f_{P_i}T_{P_i}^*f_{jP_i}}$\;\tcc{$\alpha_{P_i}^N$ is equal to $u_i$}
      \If{$N=2$}{compute $\alpha_{P_i}=\sqrt{u_i}$\;}
      \Else{compute $v_i=d_N(P_i,-P_i)\frac{f_{-P_i}}{(f_{P_i}T^*_{P_i}f_{-P_i})[-1]^*f_{P_i}}$\;\tcc{$\alpha_{P_i}^2$ is equal to $v_i$}
        compute $\alpha_{P_i}=u_i/v_i^{(N-1)/2}$\;}
 }
 \nl $S=\emptyset$\;
    \For{$i=1$ \KwTo $2g$}{
       \For{$j=2$ \KwTo $N-1$}{compute $\alpha_{jP_i}=\frac{\alpha_{(j-1)P_i}\alpha_{P_i}}{d_N((j-1)P_i,P_i)}\frac{f_{(j-1)P_i}T_{(j-1)P_i}^*f_{P_i}}{f_{jP_i}}$\;}
       \If{$S$ is not empty}{
          \For{$Q$ in $S$ and $j=1$ \KwTo $N-1$}{
          compute $\alpha_{Q+jP_i}=\frac{\alpha_Q\alpha_{jP_i}}{d_N(jP_i,Q)}\frac{f_{jP_i}T_{jP_i}^*f_{Q}}{f_{jP_i+Q}}$\;
          $S=S \cup \{Q+jP_i,jP_i\}$\;
          }
       }
       \Else{$S=\{jP_i: j=1,\ldots,N-1\}$\;}
       }
 \nl return $\{\alpha_P\}_{P\in \mathcal{A}[N]}$\;
 \caption{Compute a normalization}
\end{algorithm}

In what follows we always take $d_N$ for $N=2$ in a way like this: Let $P_1,\ldots,P_{2g}$ be a fixed symplectic basis of $\mathcal{A}[2]$. Then it takes $-1$ at the pair $(P_i,P_j)$ exactly for $j=g+i$ with $i=1,2,\ldots,g$.

\subsection{Weil sets on Weil restriction}

Let $E$ be an elliptic curve over a finite field $\mathbb{F}_{q^3}$. Let $\mathcal{W}$ be its Weil restriction with respect to the extension $\mathbb{F}_{q^3}/\mathbb{F}_q$. So $\mathcal{W}$ is an abelian
variety of dimension $3$ over $\mathbb{F}_q$, and it is isomorphic to the product $\mathcal{A}=E\times E^\sigma\times E^{\sigma^2}$ over $\mathbb{F}_{q^3}$, where $E^\sigma$ (resp. $E^{\sigma^2}$) is
obtained from $E$ by raising each coefficient of the defining equations to the $q^{th}$ (resp. $(q^2)^{th}$) power. By abuse of notation, we will use $\sigma$ for the Frobenius endomorphism on $\mathcal{W}$ and the $q^{th}$-power Frobenius maps
on $E, E^\sigma,E^{\sigma^2}$.

Let $\bar{\mathbb{F}}_q$ be the algebraic closure of $\mathbb{F}_q$. The point set $\mathcal{W}(\bar{\mathbb{F}}_q)$ can be identified with $E(\bar{\mathbb{F}}_q)\times E^\sigma(\bar{\mathbb{F}}_q) \times E^{\sigma^2}(\bar{\mathbb{F}}_q)$.
The Frobenious endomorphism $\sigma$ on $\mathcal{W}$ then takes a point $(P_1,P_2,P_3)$ to $(P_3^\sigma,P_1^\sigma,P_2^\sigma)$. So the $\mathbb{F}_q$-rational points of $\mathcal{W}$ are points of form $(P,P^\sigma,P^{\sigma^2})$ with $P\in E(\mathbb{F}_{q^3})$. The $N$-torsion subgroup of $\mathcal{W}$ corresponds to $E[N]\times E^\sigma[N] \times E^{\sigma^2}[N]$ with Weil pairing the combination of those on
$E$, $E^\sigma$ and $E^{\sigma^2}$.

Assume that $E$ is given by $y^2=x^3+a_4x+a_6$. Let $Q_i=(e_i,0)(i=1,2,3)$ be non-zero two-torsion points of $E$. Then it is easy to check that the Weil functions
\begin{eqnarray*}
f \begin{bsmallmatrix}
  0\\
  0
\end{bsmallmatrix}&=& 1, \\
 f \begin{bsmallmatrix}
  1\\
  0
\end{bsmallmatrix} &=& \sqrt{\frac{e_1-e_3}{e_1-e_2}}\frac{x-e_2}{x-e_3}, \\
 f \begin{bsmallmatrix}
  0\\
  1
\end{bsmallmatrix} &=&\sqrt{\frac{e_2-e_3}{e_2-e_1}}\frac{x-e_1}{x-e_3},\\
 f \begin{bsmallmatrix}
  1\\
  1
\end{bsmallmatrix}&=&\frac{\sqrt{(e_3-e_1)(e_2-e_3)}}{x-e_3}
\end{eqnarray*}
for $0$, $Q_1$, $Q_2$, $Q_3$ form a normal Weil set.

Fix the symplectic basis
$X_1=(Q_1,0,0)$, $X_2=(0,Q_1^\sigma,0)$, $X_3=(0,0,Q_1^{\sigma^2})$, $X_4=(Q_2,0,0)$, $X_5=(0,Q_2^\sigma,0)$, $X_6=(0,0,Q_2^{\sigma^2})$
of $\mathcal{W}[2]$. For each $(i_1,i_2,\ldots,i_6)$ in $\{0,1\}^6$, we have the Weil function
\[f\begin{bsmallmatrix}
  i_1&i_2&i_3\\
  i_4&i_5&i_6
\end{bsmallmatrix}=
f\begin{bsmallmatrix}
  i_1\\
  i_4
\end{bsmallmatrix}f\begin{bsmallmatrix}
  i_2\\
  i_5
\end{bsmallmatrix}^\sigma f\begin{bsmallmatrix}
  i_3\\
  i_6
\end{bsmallmatrix}^{\sigma^2}\] for point $\sum_{j=1}^6i_jX_j$. Here we also denote by $f\begin{bsmallmatrix}
  i\\
  j
\end{bsmallmatrix}$ (resp. $f\begin{bsmallmatrix}
  i\\
  j
\end{bsmallmatrix}^\sigma$, $f\begin{bsmallmatrix}
  i\\
  j
\end{bsmallmatrix}^{\sigma^2}$) its pullback along the projection $\mathcal{A}\to E$ (resp. $\mathcal{A}\to E^\sigma$, $\mathcal{A}\to E^{\sigma^2}$), and
$f\begin{bsmallmatrix}
  i\\
  j
\end{bsmallmatrix}^\sigma$ (resp. $f\begin{bsmallmatrix}
  i\\
  j
\end{bsmallmatrix}^{\sigma^2}$) is the function obtained from $f\begin{bsmallmatrix}
  i\\
  j
\end{bsmallmatrix}$ by raising the coefficients to the $q^{th}$ (resp. $(q^2)^{th}$) power.
As mentioned above, we take the bilinear pairing $d_2(\sum_{t=1}^6i_tX_t,\sum_{t=1}^6j_tX_t)=(-1)^{i_1j_4+i_2j_5+i_3j_6}$ on $\mathcal{W}[2]$, so that the $f\begin{bsmallmatrix}
  i_1&i_2&i_3\\
  i_4&i_5&i_6
\end{bsmallmatrix}$ form a normal Weil set, which is clear by direct computation.

\begin{remark}
 Over $\mathbb{C}$, an elliptic curve is isomorphic to a torus $\mathbb{C}/(\mathbb{Z}+\mathbb{Z}\tau)$, where $\tau$ is a complex with positive imaginary part.
 The zero loci of theta functions $\theta\begin{bsmallmatrix}
  i\\
  j
\end{bsmallmatrix}(z,\tau)=\sum_{n\in\mathbb{Z}}e^{\pi i (n+\frac{i}{2})^2\tau +2\pi in(z+\frac{j}{2})}$ with $(i,j)=(0,0),(1,0),(0,1),(1,1)$ are translations of the lattice $\mathbb{Z}+\mathbb{Z}\tau$ by $\tfrac{1+\tau}{2}$, $\tfrac{1}{2}$, $\tfrac{\tau}{2}$ and $0$ respectively, hence induce divisors on the torus. The above $f \begin{bsmallmatrix}
  i\\
  j
\end{bsmallmatrix}$ are the algebraic analogues of the ratios $(\frac{\theta\begin{bsmallmatrix}
  i\\
  j
\end{bsmallmatrix}(z,\tau)}{\theta\begin{bsmallmatrix}
  0\\
  0
\end{bsmallmatrix}(z,\tau)})^2$. The $f\begin{bsmallmatrix}
  i_1&i_2&i_3\\
  i_4&i_5&i_6
\end{bsmallmatrix}$ are constructed corresponding to the fact that if $\tau_1,\tau_2,\tau_3$ are complex numbers with positive imaginary part, then the Riemann's theta function
\[\theta\begin{bsmallmatrix}
  a\\
  b
\end{bsmallmatrix}(z,\Omega)=\sum_{n\in\mathbb{Z}^3}e^{\pi i (n+a)\Omega {}^t(n+a)+2\pi i(n+a)(z+b) }\]
satisfy
\[\theta\begin{bsmallmatrix}
  i_1&i_2&i_3\\
  i_4&i_5&i_6
\end{bsmallmatrix}((z_1,z_2,z_3),\Omega)
=\theta\begin{bsmallmatrix}
  i_1\\
  i_4
\end{bsmallmatrix}(z_1,\tau_1)\theta\begin{bsmallmatrix}
  i_2\\
  i_5
\end{bsmallmatrix}(z_2,\tau_2)\theta\begin{bsmallmatrix}
  i_3\\
  i_6
\end{bsmallmatrix}(z_3,\tau_3)\]
for the diagonal matrix $\Omega=\left(
                                  \begin{array}{ccc}
                                    \tau_1 &  &  \\
                                     & \tau_2 &  \\
                                     & &\tau_3\\
                                  \end{array}
                                \right)
$.
\end{remark}

%

In the case of $N$ being an odd prime $\ell$, we have a similar construction. Let $S_1,S_2$ be a symplectic basis of $E[\ell]$. Then
$Y_1=(S_1,0,0)$, $Y_2=(0,S_1^\sigma,0)$, $Y_3=(0,0,S_1^{\sigma^2})$, $Y_4=(S_2,0,0)$, $Y_5=(0,S_2^\sigma,0)$, $Y_6=(0,0,S_2^{\sigma^2})$ is
a symplectic basis of $\mathcal{W}[\ell]$. For $(P_1,P_2,P_3)\in \mathcal{W}[\ell]$, one can compute rational functions $f_{P_1}$, $f_{P_2}$,
$f_{P_3}$ such that
\begin{eqnarray*}
\text{div}(f_{P_1})&=&\ell(-P_1+Q_3)-\ell(Q_3)\\
\text{div}(f_{P_2})&=&\ell(-P_2+Q_3^\sigma)-\ell(Q_3^\sigma)\\
\text{div}(f_{P_3})&=&\ell(-P_3+Q_3^{\sigma^2})-\ell(Q_3^{\sigma^2}).
\end{eqnarray*}
Then $f_{P_1}f_{P_2}f_{P_3}$ is a Weil function
for $(P_1,P_2,P_3)$, that is,
\[\text{div}(f_{P_1}f_{P_2}f_{P_3})=\ell T_{(P_1,P_2,P_3)}^*\Theta-\ell\Theta\]
with $\Theta=\{Q_3\}\times E^\sigma\times E^{\sigma^2}+ E \times \{Q_3^\sigma\}\times E^{\sigma^2}+ E\times E^\sigma\times \{Q_3^{\sigma^2}\}$. To compute a normal one, we take the bilinear pairing $d_\ell$ on $\mathcal{W}[\ell]$ given by $d_\ell(P,Q)=e_\ell(P,Q)^{(\ell+1)/2}$, and apply the Algorithm \ref{algorithm_for_normalization}.

\subsection{Weil sets on Jacobian variety}
Let $C$ be a non-hyperelliptic curve of genus $3$ and $J_C$ its Jacobian variety. Fix a symplectic basis $[D_1],[D_2],\ldots,[D_6]$ of $J_C[2]$.
This basis determines a unique even theta characteristic $[\delta]$ with certain property. We want to use Weil functions corresponding to divisors $2T_P^*\Xi-2\Xi$ with $P\in J_C[2]$ and
$\Xi=\{[(R_1)+(R_2)-\delta]:R_1,R_2\in C\}$. For this,  we fix a (closed) point $O\in C$, and consider the morphism
\[j:C^3\to J_C, (R_1,R_2,R_3)\to [(R_1)+(R_2)+(R_3)-\delta-(O)].\]
It is proved in \cite[Proposition 4.1]{Barron} that the pull-backs of Weil functions along $j$ can be expressed in terms of rational functions on $C$. By abuse of notation, we denote Weil functions by
their pull-backs, which can be computed by using Algorithm \ref{WeilSetOnJV}.

\begin{algorithm}[H]\label{WeilSetOnJV}
 \KwData{a symplectic basis $[D_1],[D_2],\ldots,[D_6]$ of $J_C[2]$, a point $O\in C$}
 \KwResult{ a Weil set $\{f_{[D]}\}_{[D]\in J_C[2]}$ }
 \nl Take an effective divisor $D_0$ such that $2D_0$ is a canonical divisor on $C$\;
 \nl \For{$i=1$ \KwTo $6$}{Compute the dimension $c_i$ of the Riemann-Roch space of divisor $D_i+D_0$\;}

 \nl Take $\delta=D_0+(c_4+1)D_1+(c_5+1)D_2+(c_6+1)D_3+(c_1+1)D_4+(c_2+1)D_5+(c_3+1)D_6$\;
 \nl Compute a basis $\phi_1,\phi_2,\phi_3$ of the Riemann-Roch space of divisor $(O)+2\delta$\;
 \nl Take $f_0=1$\;
 \nl \For{$0\ne [D]\in J_C[2]$}{
      Compute a basis $\psi_1,\psi_2,\psi_3$ of the Riemann-Roch space of divisor $D+(O)+2\delta$\;
      Compute a rational function $\psi$ whose divisor is $2D$\;
      Take $f_{[D]}=(\frac{\det(\psi_i(P_j))}{\det(\phi_i(P_j))})^2\psi(P_1)\psi(P_2)\psi(P_3)$\;
   }

\nl return $\{f_{[D]}\}_{[D]\in J_C[2]}$\;
 \caption{Compute a Weil set on Jacobian variety }
\end{algorithm}

\section{Isogeny algorithm}

Let $\mathcal{W}$ be the Weil restriction of an elliptic curve $E$ over $\mathbb{F}_{q^3}$ with respect to $\mathbb{F}_{q^3}/\mathbb{F}_q$.
Let $\ell$ be an odd prime different from the characteristic of the field $\mathbb{F}_q$.
Assume that we have an $\mathbb{F}_q$-rational maximal isotropic subgroup $\Lambda$ of $\mathcal{W}[\ell]$ with respect to the Weil pairing. In order for the quotient $\mathcal{W}/\Lambda$ to be
the Jacobian variety of a curve $C$, $\Lambda$ should not contain any subgroups of $E$, $E^\sigma$ or $E^{\sigma^2}$. So for our discussion $\Lambda$ is assumed to satisfy this condition.
We denote by $\pi$ the quotient map $\mathcal{W}\to \mathcal{W}/\Lambda$.

Assume that $E$ is given by $y^2=x^3+a_4x+a_6$.  Let $Q_i=(e_i,0)(i=1,2,3)$ be non-zero two-torsion points of $E$.
Consider the divisor $\Theta=\{Q_3\}\times E^\sigma\times E^{\sigma^2}+ E \times \{Q_3^\sigma\}\times E^{\sigma^2}+ E\times E^\sigma\times \{Q_3^{\sigma^2}\}$ on $\mathcal{W}$. We already know how to
construct a normal Weil set $\{\tilde{f}_P\}_{P\in \mathcal{W}[\ell]}$, where the $\tilde{f}_P$ satisfy
\[\text{div}(\tilde{f}_P)=\ell T_P^*\Theta-\ell \Theta.\]
We remark that by construction, $\tilde{f}_{-P}=[-1]^*\tilde{f}_P$ for all $P\in \mathcal{W}[\ell]$.
By the theory of descent, the subgroup $\tilde{K}=\{(P,\tilde{f}_P):P\in \Lambda\}$ determines an invertible sheaf on $\mathcal{W}/\Lambda$ which defines a symmetric principal polarization (see for example \cite{EqnAVI}). So there is
a unique effective divisor $\Xi'$ on $\mathcal{W}/\Lambda$ such that $\pi^{-1}\Xi'-\ell\Theta$ is the divisor of a rational function $\tilde{\theta}$, which is a section of sheaf $\mathcal{O}_{\mathcal{W}}(\ell \Theta)$ invariant under the group $\tilde{K}$. We can take $\tilde{\theta}=\sum_{P\in \Lambda}\tilde{f}_P$.

Our task in this section is to give explicit equations for the curve $C$ and the homomorphism $E\to \mathcal{W}\to \mathcal{W}/\Lambda\simeq J_C$. For the equation of $C$, as in \cite{EneaMilio,lllisogeny} we use formulae in terms of analytic theta functions, which are due to Weber \cite{Weber} and Riemann \cite{Riemann}. To apply these formulae, we just need to evaluate Weil functions for two-torsion points of $\mathcal{W}/\Lambda$ at the identity element. For the homomorphism $E\to J_C$,
we first relate the Weil functions for two-torsion points of $\mathcal{W}/\Lambda$ to those for two-torsion points of $J_C$. This allows us to compute the values of Weil functions for two-torsion points of $J_C$ at the image point of a given point on $E$.
Then we describe how to recover the image point (represented by a divisor on $J_C$) from these values by solving a system of polynomial equations. If we do this for a formal point of $E$, then
we can obtain equations for the homomorphism by applying the technique of Couveignes and Ezome (see \cite{Couveignes-Ezome,EneaMilio,lllisogeny}).

\subsection{Equation for the curve $C$}

Now fix the symplectic basis
$X_1=(Q_1,0,0)$, $X_2=(0,Q_1^\sigma,0)$, $X_3=(0,0,Q_1^{\sigma^2})$, $X_4=(Q_2,0,0)$, $X_5=(0,Q_2^\sigma,0)$, $X_6=(0,0,Q_2^{\sigma^2})$
of $\mathcal{W}[2]$. Let $\bar{X}_i$ be the image of $X_i$ in $\mathcal{W}/\Lambda$. Then $\bar{X}_1, \bar{X}_2,\ldots,\bar{X}_6$ is a symplectic basis of
$\mathcal{W}/\Lambda[2]$. We want to construct Weil functions on $\mathcal{W}/\Lambda$, which correspond to divisors $2T_{\bar{P}}^*\Xi'-2\Xi'$
with $\bar{P}\in \mathcal{W}/\Lambda[2]$. Let $\{g_P\}_{P\in \mathcal{W}[2]}$ be a Weil set such that $g_P$ has divisor $2T_P^*\Theta-2\Theta$ for $P\in \mathcal{W}[2]$. Then $g_P^\ell T_P^*\tilde{\theta}^2/\tilde{\theta}^2$
has divisor $2T_{P}^*\pi^{-1}\Xi'-2\pi^{-1}\Xi'$, so we have a Weil set $\{h_{\bar{P}}\}_{\bar{P}\in \mathcal{W}/\Lambda[2]}$ such that $\pi^*h_{\bar{P}}=g_P^\ell T_P^*\tilde{\theta}^2/\tilde{\theta}^2$. Here we denote $\pi(P)$ by $\bar{P}$ for $P\in \mathcal{W}$. Let $d_2'$ be the bilinear pairing on $\mathcal{W}/\Lambda[2]$ which takes $-1$ on the $(\bar{X}_i,\bar{X}_j)$ exactly for $(i,j)=(1,4),(2,5),(3,6)$. Using $d_2'$ and Algorithm \ref{algorithm_for_normalization}, we obtain an ordered set $\{\beta_{\bar{P}}\}_{\bar{P}\in \mathcal{W}/\Lambda[2]}$ of scalars such that $\{\beta_{\bar{P}}h_{\bar{P}}\}_{\bar{P}\in \mathcal{W}/\Lambda[2]}$ is a normal Weil set.
By abuse of language, we say that $\{\beta_{\bar{P}}g_P^\ell T_P^*\tilde{\theta}^2/\tilde{\theta}^2\}_{P\in \mathcal{W}[2]}$ is a normal Weil set on $\mathcal{W}/\Lambda$.
For $(i_1,i_2,i_3,i_4,i_5,i_6)\in \{0,1\}^6$, we will denote by $h\begin{bsmallmatrix}
  i_1&i_2&i_3\\
  i_4&i_5&i_6
\end{bsmallmatrix}$ the function $\beta_{\bar{P}}g_P^\ell T_P^*\tilde{\theta}^2/\tilde{\theta}^2$ with $P=i_1X_1+i_2X_2+i_3X_3+i_4X_4+i_5X_5+i_6X_6$.

For $i=0,1,\ldots,63$, we compute
\[\vartheta_i=h\begin{bsmallmatrix}
  i_1&i_2&i_3\\
  i_4&i_5&i_6
\end{bsmallmatrix}(0),\]
where the $i_j\in\{0,1\}$ satisfy $i=i_4+2i_5+4i_6+8i_1+16i_2+32i_3$. There are two cases. If $35$ of them are non-zero, then $C$ is hyperelliptic (this can happen only if the order of $E(\mathbb{F}_{q^3})$ is even).
If $36$ of them are non-zero, then $C$ is non-hyperelliptic, and it can be represented as a plane quartic curve. We will consider only this case, since
it is bound to happen if $E$ has prime order. Now we take the Aronhold system (see \cite{Weber-formula,EneaMilio}) defined by
\begin{equation}\label{Aronhold_system}
  \begin{gathered}
  x=0, y=0,z=0,x+y+z=0,\\
  \alpha_{i1}x+\alpha_{i2}y+\alpha_{i3}z=0\, (i=1,2,3)
\end{gathered}
\end{equation}
with

\begin{equation*}
\begin{gathered}
 \alpha_{11}=1,\,\, \alpha_{21}=1, \,\, \alpha_{31}=1,\\
 \alpha_{12}=\frac{\vartheta_{5}\vartheta_{12}\vartheta_{33}\vartheta_{40}+\vartheta_{21}\vartheta_{28}\vartheta_{49}\vartheta_{56}-\vartheta_{7}\vartheta_{14}\vartheta_{35}\vartheta_{42}}{2\vartheta_{5}\vartheta_{12}\vartheta_{49}\vartheta_{56}},\\
 \alpha_{13}=\frac{\vartheta_{5}\vartheta_{12}\vartheta_{33}\vartheta_{40}+\vartheta_{7}\vartheta_{14}\vartheta_{35}\vartheta_{42}-\vartheta_{21}\vartheta_{28}\vartheta_{49}\vartheta_{56}}{2\vartheta_{5}\vartheta_{12}\vartheta_{35}\vartheta_{42}},\\
 \alpha_{22}=\frac{\vartheta_{2}\vartheta_{28}\vartheta_{47}\vartheta_{49}+\vartheta_{5}\vartheta_{27}\vartheta_{40}\vartheta_{54}-\vartheta_{14}\vartheta_{16}\vartheta_{35}\vartheta_{61}}{2\vartheta_{5}\vartheta_{27}\vartheta_{47}\vartheta_{49}},\\
\alpha_{23}=\frac{\vartheta_{2}\vartheta_{28}\vartheta_{47}\vartheta_{49}-\vartheta_{5}\vartheta_{27}\vartheta_{40}\vartheta_{54}-\vartheta_{14}\vartheta_{16}\vartheta_{35}\vartheta_{61}}{2\vartheta_{5}\vartheta_{27}\vartheta_{35}\vartheta_{61}},\\
\alpha_{32}=\frac{\vartheta_{7}\vartheta_{16}\vartheta_{42}\vartheta_{61}-\vartheta_{12}\vartheta_{27}\vartheta_{33}\vartheta_{54}-\vartheta_{2}\vartheta_{21}\vartheta_{47}\vartheta_{56}}{-2\vartheta_{12}\vartheta_{27}\vartheta_{47}\vartheta_{56}},\\
\alpha_{33}=\frac{\vartheta_{12}\vartheta_{27}\vartheta_{33}\vartheta_{54}+\vartheta_{7}\vartheta_{16}\vartheta_{42}\vartheta_{61}-\vartheta_{2}\vartheta_{21}\vartheta_{47}\vartheta_{56}}{-2\vartheta_{12}\vartheta_{27}\vartheta_{42}\vartheta_{61}}.
\end{gathered}
\end{equation*}
By Riemann's theorem (see \cite[Proposition 3]{Ritzenthaler}), we first compute $u_1, u_2, u_3$, $k_1, k_2, k_3$ such that
\[\left(\begin{matrix}
       \frac{1}{\alpha_{11}} & \frac{1}{\alpha_{21}} & \frac{1}{\alpha_{31}} \\
     \frac{1}{\alpha_{12}} & \frac{1}{\alpha_{22}} & \frac{1}{\alpha_{32}} \\
      \frac{1}{\alpha_{13}} & \frac{1}{\alpha_{23}} & \frac{1}{\alpha_{33}}\\
    \end{matrix}\right)
  \left(
           \begin{matrix}
             u_1 \\
             u_2 \\
             u_3 \\
           \end{matrix}
         \right)\!=\!\left(
                   \begin{matrix}
                     -1 \\
                     -1 \\
                    -1 \\
                   \end{matrix}
                 \right),
                 \left(
                   \begin{matrix}
                      u_1\alpha_{11} & u_2\alpha_{21} & u_3\alpha_{31} \\
                      u_1\alpha_{12} & u_2\alpha_{22} & u_3\alpha_{32} \\
                      u_1\alpha_{13} & u_2\alpha_{23} & u_3\alpha_{33}\\
                   \end{matrix}
                 \right)\left(
                          \begin{matrix}
                            k_1 \\
                            k_2 \\
                            k_3 \\
                          \end{matrix}
                        \right)\!=\!\left(
                                  \begin{matrix}
                                    -1 \\
                                    -1 \\
                                    -1 \\
                                  \end{matrix}
                                \right).
\]
Next, we compute the linear functions $\xi_1,\xi_2,\xi_3$ of $x,y,z$ which are determined by the equation
\[\left(
    \begin{matrix}
      1 & 1 & 1 \\
      \frac{1}{\alpha_{11}} & \frac{1}{\alpha_{12}} & \frac{1}{\alpha_{13}} \\
     \frac{1}{\alpha_{21}} & \frac{1}{\alpha_{22}} & \frac{1}{\alpha_{23}} \\
      \frac{1}{\alpha_{31}} & \frac{1}{\alpha_{32}} & \frac{1}{\alpha_{33}}\\
    \end{matrix}
  \right)\left(
           \begin{matrix}
             \xi_1 \\
             \xi_2 \\
             \xi_3\\
           \end{matrix}
         \right)=-\left(
    \begin{matrix}
      1 & 1 & 1 \\
      k_1\alpha_{11} & k_1\alpha_{12} & k_1\alpha_{13} \\
     k_2\alpha_{21} & k_2\alpha_{22} & k_2\alpha_{23} \\
      k_3\alpha_{31} & k_3\alpha_{32} & k_3\alpha_{33}\\
    \end{matrix}
  \right)\left(
           \begin{matrix}
             x \\
             y \\
             z\\
           \end{matrix}
         \right).
\]
Then $(x\xi_1+y\xi_2-z\xi_3)^2=4xy\xi_1\xi_2$ is an equation of the curve $C$.

Note that the coefficients of the above equation for $C$ are in an extension field of $\mathbb{F}_q$.
We can compute its normalised Dixmier-Ohno invariants and construct a new equation with coefficients
in $\mathbb{F}_q$ and same invariants \cite{Ritzenthaler_quartics}.
We then check whether $\mathcal{W}/\Lambda$, which is a Jacobian over $\mathbb{F}_{q^2}$, is the Jacobian of the curve over $\mathbb{F}_q$ defined by the new equation.
This can be done by checking that the image of a randomly chosen $\mathbb{F}_q$-rational point in the Jacobian under the multiplication-by-$\#E(\mathbb{F}_{q^3})$ map is the identity element.
Now we assume that $\mathcal{W}/\Lambda$ is a Jacobian over $\mathbb{F}_q$. Because all two-torsion points of $J_C$ are $\mathbb{F}_{q^9}$-rational, there must be an $\mathbb{F}_q$-rational bitangent.
After a suitable linear change of coordinates in
$\mathbb{P}^2$, we may assume that $z=0$ is a bitangent of the curve over $\mathbb{F}_q$.
Then it is easy to compute a linear change of coordinates in $\mathbb{P}^2$ which induces an isomorphism between the curve defined by $(x\xi_1+y\xi_2-z\xi_3)^2=4xy\xi_1\xi_2$
and the curve over $\mathbb{F}_q$. So we will think of $C$ as a curve over $\mathbb{F}_q$, together with the Aronhold system obtained by applying the corresponding linear change of coordinates to (\ref{Aronhold_system}).

\subsection{Computation of image point}

Let $L_1,\ldots,L_7$ be the odd theta characteristic determined by the bitangents in the Aronhold system. Then we compute an even theta characteristic $[\delta]$ and a symplectic basis
$[D_1],\ldots, [D_6]$ such that
\begin{equation}\label{theta_character_symplectic_basis}
(L_1-[\delta],\ldots, L_7-[\delta])=([D_1],\ldots,[D_6])\begin{bmatrix}
1&0&0&1&1&1&0\\
1&0&1&0&0&1&1\\
1&1&1&1&0&0&0\\
1&0&0&1&1&0&1\\
1&1&0&0&0&1&1\\
1&1&1&0&1&0&0
\end{bmatrix}.
\end{equation}

Let $\Xi$ be the theta divisor $\{[(R_1)+(R_2)-\delta]: R_1,R_2\in C\}$ on $J_C$. Let $K_C=2(\infty_1)+2(\infty_2)$ be the canonical divisor on $C$ cut by $z=0$. By using Algorithm \ref{WeilSetOnJV} with
the above symplectic basis $[D_1],\ldots, [D_6]$ and point $\infty_1$ as input, we obtain a Weil set on $J_C[2]$. Let $d_2''$ be the bilinear pairing on $J_C[2]$ which takes $-1$ on $([D_i],[D_j])$ exactly for $(i,j)=(1,4),(2,5),(3,6)$. Then we use $d_2''$ to transform it into a normal one. As before, we denote by $\xi\begin{bsmallmatrix}
  i_1&i_2&i_3\\
  i_4&i_5&i_6
\end{bsmallmatrix}$ the resulting Weil function for point $i_1[D_1]+\cdots+i_6[D_6]$.

We may now ask what is the relation between $h\begin{bsmallmatrix}
  i_1&i_2&i_3\\
  i_4&i_5&i_6
\end{bsmallmatrix}$ and $\xi\begin{bsmallmatrix}
  i_1&i_2&i_3\\
  i_4&i_5&i_6
\end{bsmallmatrix}$.
Recall that the $h\begin{bsmallmatrix}
  i_1&i_2&i_3\\
  i_4&i_5&i_6
\end{bsmallmatrix}$ are the pull-backs of Weil functions by $\mathcal{W}\to \mathcal{W}/\Lambda$, and that
 the $\xi\begin{bsmallmatrix}
  i_1&i_2&i_3\\
  i_4&i_5&i_6
\end{bsmallmatrix}$ are rational functions on $C^3$ which are the pull-backs of Weil functions by the morphism
\[j:C^3\to J_C,(R_1,R_2,R_3)\mapsto [(R_1)+(R_2)+(R_3)-\delta-(\infty_1)].\]
Now define
  \begin{align*}
    \chi_1&=\frac{h\begin{bsmallmatrix}
  1&0&0\\
  0&0&0
\end{bsmallmatrix}(0)}{\xi\begin{bsmallmatrix}
  1&0&0\\
  0&0&0
\end{bsmallmatrix}(0)},  &\chi_2=&\frac{h\begin{bsmallmatrix}
  0&1&0\\
  0&0&0
\end{bsmallmatrix}(0)}{\xi\begin{bsmallmatrix}
  0&1&0\\
  0&0&0
\end{bsmallmatrix}(0)}, &\chi_3=&\frac{h\begin{bsmallmatrix}
  0&0&1\\
  0&0&0
\end{bsmallmatrix}(0)}{\xi\begin{bsmallmatrix}
  0&0&1\\
  0&0&0
\end{bsmallmatrix}(0)},   \\
    \chi_4&=\frac{h\begin{bsmallmatrix}
  0&0&0\\
  1&0&0
\end{bsmallmatrix}(0)}{\xi\begin{bsmallmatrix}
  0&0&0\\
  1&0&0
\end{bsmallmatrix}(0)},  &\chi_5=&\frac{h\begin{bsmallmatrix}
  0&0&0\\
  0&1&0
\end{bsmallmatrix}(0)}{\xi\begin{bsmallmatrix}
  0&0&0\\
  0&1&0
\end{bsmallmatrix}(0)}, &\chi_6=&\frac{h\begin{bsmallmatrix}
  0&0&0\\
  0&0&1
\end{bsmallmatrix}(0)}{\xi\begin{bsmallmatrix}
  0&0&0\\
  0&0&1
\end{bsmallmatrix}(0)}.
  \end{align*}
Here the $\xi\begin{bsmallmatrix}
  i_1&i_2&i_3\\
  i_4&i_5&i_6
\end{bsmallmatrix}(0)$ mean that the functions are regarded as functions on $J_C$, and evaluated at the identity element. This is possible since the identity element is not in $\Xi$ and the functions
are regular at $j^{-1}(0)$. The $h\begin{bsmallmatrix}
  i_1&i_2&i_3\\
  i_4&i_5&i_6
\end{bsmallmatrix}(0)$ are understood in the same way. The relation then can be expressed as follows. Let $P$ be a point of $\mathcal{W}(\bar{\mathbb{F}}_q)$ and $(R_1,R_2,R_3)$ the point of $C^3(\bar{\mathbb{F}}_q)$ such that $j(R_1,R_2,R_3)$
is the image of $P$ under the homomorphism $\mathcal{W}\to \mathcal{W}/\Lambda\simeq J_C$. Then for $(i_1,\ldots,i_6)\in\{0,1\}^6$, we have
\begin{equation}\label{relation}
  \xi\begin{bsmallmatrix}
  i_1&i_2&i_3\\
  i_4&i_5&i_6
\end{bsmallmatrix}(R_1,R_2,R_3)=\chi_1^{i_1}\chi_2^{i_2}\cdots\chi_6^{i_6}h\begin{bsmallmatrix}
  i_1&i_2&i_3\\
  i_4&i_5&i_6
\end{bsmallmatrix}(P).
\end{equation}

It remains to compute $(R_1,R_2,R_3)$ from the values $\xi\begin{bsmallmatrix}
  i_1&i_2&i_3\\
  i_4&i_5&i_6
\end{bsmallmatrix}(R_1,R_2,R_3)$. Since it seems difficult to solve the system of equations directly, we do it another way.
The idea is to replace the $\xi\begin{bsmallmatrix}
  i_1&i_2&i_3\\
  i_4&i_5&i_6
\end{bsmallmatrix}$ by some "good" functions which allow us to construct a system of polynomial equations of low degree.

Let $x_1, y_1$, $x_2, y_2$, $x_3, y_3$ be the affine coordinate functions on the factors of $C^3$. Then the functions $f$ with $\text{div}(f)+2\Xi\geq 0$
satisfy
\[\text{div}(\omega^2j^*f)+\sum_{i=1}^3pr_i^*(2(\infty_1)+2K_C)\geq 2\Delta,\]
where $\omega=\det\begin{bsmallmatrix}
      x_1 & y_1 & 1 \\
      x_2 & y_2 & 1 \\
      x_3 & y_3 & 1 \\
    \end{bsmallmatrix}$, $\Delta$ is the full diagonal of $C^3$. We want to find eight functions $\zeta_1,\ldots,\zeta_8$ on $C^3$ such that $\omega^2\xi\begin{bsmallmatrix}
  i_1&i_2&i_3\\
  i_4&i_5&i_6
\end{bsmallmatrix}$ is
a linear combination of $\zeta_1,\ldots,\zeta_8$. Since the Riemann-Roch space of the divisor $2(\infty_1)+2K_C$ has a basis $\mathfrak{B}=\{1,x,x^2,y,y^2,xy,u,v\}$ with $u,v$ polynomials in $x,y$ of degree $3$,
we consider the functions
\begin{equation}\label{symmetric}
\sum_{\tau\in \text{Sym}(\{1,2,3\})}pr_{\tau(1)}^*t_1 \cdot pr_{\tau(2)}^*t_2\cdot  pr_{\tau(3)}^*t_3 \,\,(t_1,t_2,t_3\in\mathfrak{B}).
\end{equation}
There are $120$ different functions. We pick $n\ge 120$ points in $\Delta$, and compute the values of these functions at each point to get an $n\times 120$ matrix. If $n$ is large enough, this matrix has rank $112$, and we
can deduce $8$ linearly independent functions $\zeta_1,\ldots,\zeta_8$.

We compute a maximal linearly independent subset of the $\xi\begin{bsmallmatrix}
  i_1&i_2&i_3\\
  i_4&i_5&i_6
\end{bsmallmatrix}$. For simplicity of notation, we assume that the $\xi\begin{bsmallmatrix}
  i_1&i_2&i_3\\
  0&0&0
\end{bsmallmatrix}$ are linearly independent. Then by evaluating the $16$ functions at $9$ points of $C^3$ in general we can find the scalars $c_{i_1,i_2,i_3}^{(i)}$ such that
\[\zeta_i=\sum_{(i_1,i_2,i_3)\in\{0,1\}^3}c_{i_1,i_2,i_3}^{(i)} \omega^2\xi\begin{bsmallmatrix}
  i_1&i_2&i_3\\
  0&0&0
\end{bsmallmatrix} (i=1,2,\ldots,8).\]

Now for $P\in \mathcal{W}(\bar{\mathbb{F}}_q)$, we can compute
\[\lambda_i= \sum_{(i_1,i_2,i_3)\in\{0,1\}^3}c_{i_1,i_2,i_3}^{(i)} \chi_1^{i_1}\chi_2^{i_2}\chi_3^{i_3}h\begin{bsmallmatrix}
  i_1&i_2&i_3\\
  0&0&0
\end{bsmallmatrix}(P)\]
for $i=1,2,\ldots,8$. Assume that $R_i=(a_i:b_i:1)$. Then we can compute them by solving the system defined by
\begin{gather*}
 C(a_i,b_i,1)=0 (i=1,2,3),\,  \lambda \zeta_{i_0}(a_1,b_1,a_2,b_2,a_3,b_3)=1, \\
  \zeta_1(a_1,b_1,a_2,b_2,a_3,b_3)\lambda_i=\zeta_i(a_1,b_1,a_2,b_2,a_3,b_3)\lambda_1(i=2,3,\ldots,8),
\end{gather*}
where $C$ denotes also the equation of $C$, $\zeta_{i_0}(R_1,R_2,R_3)$ is assumed to be non-zero for a fixed $i_0\in \{1,2,\ldots,8\}$, and $a_1,b_1,a_2,b_2,a_3,b_3,\lambda$ are indeterminates.
This system is in practice easy to solve by computing the Gr\"obner basis of the corresponding ideal. The results are two opposite points on $J_C$, that is,
there is another point $(\bar{R}_1,\bar{R}_2,\bar{R}_3)\in C^3$ such that $j(R_1,R_2,R_3)=-j(\bar{R}_1,\bar{R}_2,\bar{R}_3)$ and that $\zeta_i(R_1,R_2,R_3)=\zeta_i(\bar{R}_1,\bar{R}_2,\bar{R}_3)$
for $i=1,2,\ldots,8$.

\subsection{Equations for the homomorphism}
The results in previous subsection is sufficient to translate the discrete logarithm problem from $E(\mathbb{F}_{q^3})$ to $J_C(\mathbb{F}_q)$. In \cite{Couveignes-Ezome}, Couveignes and Ezome
go far beyond computing the image of a point under an $(\ell,\ell)$-isogeny between Jacobians of genus $2$ curves. They construct an explicit map from the curve to the Jacobian by solving certain system of differential equations.
Their idea can be adapted to give a description of our homomorphism (see also \cite{lllisogeny}).

Let $\varphi: E\to J_C$ be the composition of the map
\[i: E\to \mathcal{W}, P\mapsto (P,0,0)\]
and the quotient map $\pi:\mathcal{W}\to \mathcal{W}/\Lambda=J_C$. Given $P_0\in E(\mathbb{F}_{q^3})$ with $Q_0\in \langle P_0 \rangle$, the discrete logarithm of $Q_0$ with respect to $P_0$ is the
discrete logarithm of $\varphi(Q_0)+\varphi(Q_0)^\sigma+\varphi(Q_0)^{\sigma^2}$ with respect to $\varphi(P_0)+\varphi(P_0)^\sigma+\varphi(P_0)^{\sigma^2}\in J_C(\mathbb{F}_q)$.
We will give an expression for $\varphi$ instead of the map $P\mapsto \pi(P,P^\sigma,P^{\sigma^2})=\varphi(P)+\varphi(P)^\sigma+\varphi(P)^{\sigma^2}$.

For $P=(x_0,y_0)\in E$, there are points $R_1,R_2,R_3\in C$ such that $\varphi(P)$ is represented by the class of divisor $(R_1)+(R_2)+(R_3)-\delta-(\infty_1)$. We consider the following functions:
\begin{eqnarray*}
   \mathbf{C}_1(x_0,y_0)&=& x(R_1)+x(R_2)+x(R_3),\\
   \mathbf{C}_2(x_0,y_0)&=& x(R_1)x(R_2)+x(R_1)x(R_3) +x(R_2)x(R_3),\\
   \mathbf{C}_3(x_0,y_0)&=& x(R_1)x(R_2)x(R_3),\\
   \mathbf{C}_4(x_0,y_0)&=& y(R_1)+y(R_2)+y(R_3),\\
   \mathbf{C}_5(x_0,y_0)&=& y(R_1)y(R_2)+y(R_1)y(R_3) +y(R_2)y(R_3),\\
   \mathbf{C}_6(x_0,y_0)&=& y(R_1)y(R_2)y(R_3).
\end{eqnarray*}
The effective divisor $(R_1)+(R_2)+(R_3)$ (in general) is the divisor cut by
\[x^3-\mathbf{C}_1(x_0,y_0)x^2z+\mathbf{C}_2(x_0,y_0)xz^2-\mathbf{C}_3(x_0,y_0)z^3\]
  and
\[y^3-\mathbf{C}_4(x_0,y_0)y^2z+\mathbf{C}_5(x_0,y_0)yz^2-\mathbf{C}_6(x_0,y_0)z^3.\]
The $\mathbf{C}_i(x,y)=\mathbf{A}_i(x)+\mathbf{B}_i(x)y$ are rational functions on $E$. If $-\varphi(P)=[(\bar{R}_1)+(\bar{R}_2)+(\bar{R}_3)-\delta-(\infty_1)]$, then
\begin{eqnarray*}
  \mathbf{C}_1(x_0,-y_0)&=&x(\bar{R}_1)+x(\bar{R}_2)+x(\bar{R}_3), \\
  &\vdots&\\
  \mathbf{C}_6(x_0,-y_0)&=& y(\bar{R}_1)y(\bar{R}_2)y(\bar{R}_3).
\end{eqnarray*}

We denote by $C(x,y,z)$ the defining polynomial of the plane quartic $C$. Let $C_y(x,y)$ be the derivative of $C(x,y,1)$ with respect to the variable $y$. Then $\frac{dx}{C_y(x,y)}$, $\frac{xdx}{C_y(x,y)}$, $\frac{ydx}{C_y(x,y)}$
is a basis of the vector space $H^0(C,\Omega_C^1)$ of regular differentials on $C$. The vector space $H^0(J_C,\Omega_{J_C}^1)$ of regular differentials on $J_C$ can be identified with the invariant subspace of
$H^0(C^3,\Omega_{C^3}^1)=pr_1^*H^0(C,\Omega_C^1)\oplus pr_2^*H^0(C,\Omega_C^1)\oplus pr_3^*H^0(C,\Omega_C^1)$ by permutations of the factors. This space has a basis
\begin{eqnarray*}
\omega_1 &=& \frac{dx_1}{C_y(x_1,y_1)}+\frac{dx_2}{C_y(x_2,y_2)}+\frac{dx_3}{C_y(x_3,y_3)}, \\
\omega_2 &=& \frac{x_1dx_1}{C_y(x_1,y_1)}+\frac{x_2dx_2}{C_y(x_2,y_2)}+\frac{x_3dx_3}{C_y(x_3,y_3)},\\
\omega_3&=&\frac{y_1dx_1}{C_y(x_1,y_1)}+\frac{y_2dx_2}{C_y(x_2,y_2)}+\frac{y_3dx_3}{C_y(x_3,y_3)}.
\end{eqnarray*}
The pull-backs of $\omega_1,\omega_2,\omega_3$ under $\varphi$ are regular, hence there are constants $m_1,m_2,m_3$ such that $\varphi^*\omega_i=m_i\frac{dx}{y}$ for $i=1,2,3$.

Now let $t=x-x_0$ be the local parameter at $P$. We compute a formal point $P(t)=(\mu(t),\nu(t))$ with $\mu(t)=x_0+t$ and $\nu(0)=y_0\ne 0$.
Let $R_i(t)=(\alpha_i(t),\beta_i(t))$, $\bar{R}_i(t)=(\bar{\alpha}_i(t),\bar{\beta}_i(t))$, $i=1,2,3$ be formal points on $C$ such that
\begin{eqnarray*}
 \varphi(P(t))&=&[(R_1(t))+(R_2(t))+(R_3(t))-(\infty_1)-\delta],  \\
 \varphi(\bar{P}(t))&=&[(\bar{R}_1(t))+(\bar{R}_2(t))+(\bar{R}_3(t))-(\infty_1)-\delta]
\end{eqnarray*}
with $\bar{P}(t)=(\mu(t), -\nu(t))$.
 These formal points satisfy the following systems of differential equations
\begin{equation}\label{differential_eqn1}
  \left\{ \begin{aligned}
   \frac{\dot{\alpha}_1(t)}{C_{y}(\alpha_1(t),\beta_1(t))}+\frac{\dot{\alpha}_2(t)}{C_{y}(\alpha_2(t),\beta_2(t))}+\frac{\dot{\alpha}_3(t)}{C_{y}(\alpha_3(t),\beta_3(t))}&=\frac{m_1\dot{\mu}(t)}{\nu(t)},\\
   \frac{\alpha_1(t)\dot{\alpha}_1(t)}{C_{y}(\alpha_1(t),\beta_1(t))}+\frac{\alpha_2(t)\dot{\alpha}_2(t)}{C_{y}(\alpha_2(t),\beta_2(t))}+\frac{\alpha_3(t)\dot{\alpha}_3(t)}{C_{y}(\alpha_3(t),\beta_3(t))}&=\frac{m_2\dot{\mu}(t)}{\nu(t)},\\
   \frac{\beta_1(t)\dot{\alpha}_1(t)}{C_{y}(\alpha_1(t),\beta_1(t))}+\frac{\beta_2(t)\dot{\alpha}_2(t)}{C_{y}(\alpha_2(t),\beta_2(t))}+\frac{\beta_3(t)\dot{\alpha}_3(t)}{C_{y}(\alpha_3(t),\beta_3(t))}&=\frac{m_3\dot{\mu}(t)}{\nu(t)},\\
  C(\alpha_1(t),\beta_1(t))&=0,\\
   C(\alpha_2(t),\beta_2(t))&=0,\\
   C(\alpha_3(t),\beta_3(t))&=0.
   \end{aligned}\right.
\end{equation}
and
\begin{equation}\label{differential_eqn2}
  \left\{ \begin{aligned}
   \frac{\dot{\bar{\alpha}}_1(t)}{C_{y}(\bar{\alpha}_1(t),\bar{\beta}_1(t))}+\frac{\dot{\bar{\alpha}}_2(t)}{C_{y}(\bar{\alpha}_2(t),\bar{\beta}_2(t))}+\frac{\dot{\bar{\alpha}}_3(t)}{C_{y}(\bar{\alpha}_3(t),\bar{\beta}_3(t))}&=\frac{\bar{m}_1\dot{\mu}(t)}{\nu(t)},\\
   \frac{\bar{\alpha}_1(t)\dot{\bar{\alpha}}_1(t)}{C_{y}(\bar{\alpha}_1(t),\bar{\beta}_1(t))}+\frac{\bar{\alpha}_2(t)\dot{\bar{\alpha}}_2(t)}{C_{y}(\bar{\alpha}_2(t),\bar{\beta}_2(t))}+\frac{\bar{\alpha}_3(t)\dot{\bar{\alpha}}_3(t)}{C_{y}(\bar{\alpha}_3(t),\bar{\beta}_3(t))}&=\frac{\bar{m}_2\dot{\mu}(t)}{\nu(t)},\\
   \frac{\beta_1(t)\dot{\bar{\alpha}}_1(t)}{C_{y}(\bar{\alpha}_1(t),\bar{\beta}_1(t))}+\frac{\bar{\beta}_2(t)\dot{\bar{\alpha}}_2(t)}{C_{y}(\bar{\alpha}_2(t),\bar{\beta}_2(t))}+\frac{\bar{\beta}_3(t)\dot{\bar{\alpha}}_3(t)}{C_{y}(\bar{\alpha}_3(t),\bar{\beta}_3(t))}&=\frac{\bar{m}_3\dot{\mu}(t)}{\nu(t)},\\
  C(\bar{\alpha}_1(t),\bar{\beta}_1(t))&=0,\\
   C(\bar{\alpha}_2(t),\bar{\beta}_2(t))&=0,\\
   C(\bar{\alpha}_3(t),\bar{\beta}_3(t))&=0.
   \end{aligned}\right.
\end{equation}
The differential equations correspond to the fact that we can express the pull-backs of differentials $\omega_1,\omega_2,\omega_3$ under $\varphi$ (and $\varphi\circ [-1]$)
in terms of the differential $\frac{dx}{y}$ on $E$.

To compute the $\mathbf{A}_i, \mathbf{B}_i$, we proceed as follows.
\begin{enumerate}
  \item Compute the $R_i(t)$ (resp. $\bar{R}_i(t)$) at precision $3$ with the method in previous subsection. By comparing the coefficient of $t^0$ in (\ref{differential_eqn1}) (resp. (\ref{differential_eqn2})), we obtain the values $m_1,m_2,m_3$ (resp. $\bar{m}_1,\bar{m}_2,\bar{m}_3$).
  \item Increase the accuracy of the $\alpha_i(t),\beta_i(t)$ (resp. $\bar{\alpha}_i(t),\bar{\beta}_i(t)$). Their coefficients can be computed one by one by using the system (\ref{differential_eqn1}) (resp. (\ref{differential_eqn2})).
  \item Compute
            \begin{eqnarray*}
             \mathbf{A}_1(\mu(t))&=&\frac{\alpha_1(t)+\alpha_2(t)+\alpha_3(t)+\bar{\alpha}_1(t)+\bar{\alpha}_2(t)+\bar{\alpha}_3(t)}{2},  \\
             \mathbf{B}_1(\mu(t))&=& \frac{\alpha_1(t)+\alpha_2(t)+\alpha_3(t)-\bar{\alpha}_1(t)-\bar{\alpha}_2(t)-\bar{\alpha}_3(t)}{2\nu(t)},\\
             &\vdots&\\
             \mathbf{A}_6(\mu(t))&=&\frac{\beta_1(t)\beta_2(t)\beta_3(t)+\bar{\beta}_1(t)\bar{\beta}_2(t)\bar{\beta}_3(t)}{2},  \\
             \mathbf{B}_6(\mu(t))&=& \frac{\beta_1(t)\beta_2(t)\beta_3(t)-\bar{\beta}_1(t)\bar{\beta}_2(t)\bar{\beta}_3(t)}{2\nu(t)}.\\
            \end{eqnarray*}
  Then recover the $\mathbf{A}_i, \mathbf{B}_i$ by using continued fraction.
\end{enumerate}

\subsection{An example}
We give an example computed with the computational algebra system Magma \cite{magma}.

Let $\mathbb{F}_{523^3}=\mathbb{F}_{523}[w]$ be the cubic extension of $\mathbb{F}_{523}$ defined by $w^3+2=0$, and let
$E$ be the elliptic curve over $\mathbb{F}_{523^3}$ defined by
\[y^2=x^3+(82w^2 + 92w + 140)x+359w^2 + 339w + 243.\]
Then the order of the group $E(\mathbb{F}_{523^3})$ is the prime $143069629$.

Let $\mathcal{W}$ be the Weil restriction of $E$ with respect to $\mathbb{F}_{523^3}/\mathbb{F}_{523}$. Let $\mathbb{F}_{523^{6}}=\mathbb{F}_{523^3}[\theta]$ be given by $\theta^2 + 522w + 517=0$.
Then we take the $\mathbb{F}_{523}$-rational subgroup $\Lambda=\langle P_1,P_2,P_3\rangle$ of $\mathcal{W}[3]$ generated by
\begin{eqnarray*}
P_1&=&\bigl(\,(222w^2+370w+337,(122w^2+387w+400)\theta),\\
&&\quad (312w^2+287w+156,(21w^2+362w+493)\theta),\\
&&\quad (415w^2+275w + 156,(419w^2 + 251w + 451)\theta)\bigr), \\
P_2&=&\bigl((319w^2 + 484w + 156, (298w^2 + 484w + 108)\theta),\\
&&\quad (245w^2 + 442w + 337,(305w^2 + 456w + 6)\theta),\\
&& \quad (415w^2 +275w + 156,(104w^2 + 272w + 72)\theta)\bigr),\\
P_3&=&\bigl((319w^2 + 484w + 156,(225w^2 + 39w + 415)\theta),\\
&&\quad  (312w^2 + 287w + 156 ,(21w^2 + 362w + 493)\theta), \\
&& \quad (56w^2 + 234w + 337 , (150w^2 + 109w +222)\theta)\bigr).
\end{eqnarray*}
The quotient $\mathcal{W}/\Lambda$ is isomorphic to the Jacobian of curve $C$ defined by
\begin{multline*}
  x^4 + 144x^3y + 112x^2y^2 + 393xy^3 + 488y^4 + 397x^3z + 49x^2yz + 299xy^2z + 500y^3z\\
+ 311x^2z^2 +185xyz^2 + 288y^2z^2 + 462xz^3 + 468yz^3 + 284z^4=0.
\end{multline*}

Let $\varphi: E\to J_C$ be the composition of the map
\[i: E\to \mathcal{W}, P\mapsto (P,0,0)\]
and the quotient map $\pi:\mathcal{W}\to \mathcal{W}/\Lambda=J_C$. If $P=(x_0,y_0)\in E$, then
$\varphi(P)$ is the class of the divisor $D-3(\infty_1)$ with $\infty_1=(359:1:0)$ and $D$
cut by
\begin{align*}
  & x^3-\mathbf{C}_1(x_0,y_0)x^2z+\mathbf{C}_2(x_0,y_0)xz^2-\mathbf{C}_3(x_0,y_0)z^3, \\
  & y^3-\mathbf{C}_4(x_0,y_0)y^2z+\mathbf{C}_5(x_0,y_0)yz^2-\mathbf{C}_6(x_0,y_0)z^3,
\end{align*}
where $\mathbf{C}_1, \mathbf{C}_2,\ldots,\mathbf{C}_6$ are rational functions on $E$ given by
\begin{align*}
\mathbf{C}_1(x,y)&=\Bigl(\bigl((407w^2 + 492w + 158)x^{10} + (90w^2 + 63w + 215)x^9 + (141w^2 + 410w + 456)x^8\\
     & + (118w^2 + 156w + 463)x^7 +(417w^2 + 319w + 470)x^6 + (481w^2 + 406w + 109)x^5\\
     & + (176w^2 + 194w + 77)x^4 + (337w^2 + 298w + 503)x^3 +(167w^2 + 458w + 409)x^2\\
     & + (456w^2 + 163w + 256)x + 12w^2 + 203w + 42\bigr)+ y\bigl((181w^2 + 196w + 325)x^8 \\
     &+ (358w^2 + 100w + 386)x^7 + (497w^2 + 485w + 205)x^6 + (453w^2 + 238w + 134)x^5 \\
     &+ (223w^2 + 289w + 381)x^4 + (195w^2 + 74w + 88)x^3 + (414w^2 + 200w + 57)x^2 \\
     &+ (153w^2 + 135w + 428)x + 7w^2+ 325w + 98\bigr)\Bigr)/\mathbf{p}(x),\\
\end{align*}

\begin{align*}
\mathbf{C}_2(x,y)&=\Bigl(\bigl((228w^2 + 138w + 383)x^{10} + (161w^2 + 421w + 392)x^9 + (194w^2 + 522w + 351)x^8\\
& + (261w^2 + 269w + 190)x^7 + (504w^2 + 72w + 309)x^6 + (194w^2 + 249w + 4)x^5\\
& + (247w^2 + 179w + 463)x^4 + (292w^2 + 158w + 298)x^3 +(67w^2 + 488w + 196)x^2\\
& + (379w^2 + 336w + 513)x + 257w^2 + 103w + 125\bigr)+y\bigl((456w^2 + 509w + 243)x^9 \\
&+ (89w^2 + 121w + 21)x^8 + (107w^2 + 175w + 165)x^7 + (437w^2 + 50w + 473)x^6\\
& +(448w^2 + 139w + 447)x^5 + (491w^2 + 324w + 384)x^4 + (38w^2 + 272w + 229)x^3 \\
&+ (107w^2 + 95w + 314)x^2 +(200w^2 + 143w + 122)x + 221w^2 + 128w + 162\bigr)\Bigr)/\mathbf{p}(x),
\end{align*}

\begin{align*}
\mathbf{C}_3(x,y)&=\Bigl(\bigl((326w^2 + 484w + 272)x^{11} + (376w^2 + 332w + 245)x^{10} + (239w^2 + 145w + 95)x^9\\
& + (106w^2 + 106w + 233)x^8+ (184w^2 + 418w + 403)x^7 + (240w^2 + 220w + 239)x^6\\
& + (127w^2 + 286w + 468)x^5 + (484w^2 + 516w + 68)x^4 +(170w^2 + 335w + 124)x^3\\
& + (462w^2 + 426w + 433)x^2 + (98w^2 + 500w + 460)x + 509w^2 + 135w + 101\bigr)\\
&+y\bigl((371w^2 + 242w + 329)x^9 + (68w^2 + 87w + 240)x^8 + (21w^2 + 389w + 116)x^7 \\
&+ (110w^2 + 79w + 107)x^6 +(212w^2 + 190w + 216)x^5 + (332w^2 + 382w + 182)x^4\\
& + (93w^2 + 33w + 465)x^3+ (280w^2 + 192w + 124)x^2 +(395w^2 + 463w + 192)x \\
&+ 112w^2 + 170w + 292\bigr)\Bigr)/\mathbf{p}(x),
\end{align*}

\begin{align*}
\mathbf{C}_4(x,y)&=\Bigl(\bigl((513w^2 + 260w + 140)x^{10} + (347w^2 + 92w + 344)x^9 + (314w^2 + 289w + 405)x^8\\
& + (199w^2 + 377w + 117)x^7 +(153w^2 + 414w + 96)x^6 + (507w^2 + 419w + 349)x^5\\
& + (20w^2 + 289w + 74)x^4 + (36w^2 + 407w + 235)x^3 +(299w^2 + 496w + 429)x^2 \\
&+ (5w^2 + 166w + 6)x + 63w^2 + 13w + 356\bigr)+y\bigl((20w^2 + 462w + 217)x^8 \\
&+ (41w^2 + 359w + 88)x^7 + (370w^2 + 137w + 358)x^6 + (470w^2 + 473w + 21)x^5 \\
&+(103w^2 + 479w + 1)x^4 + (270w^2 + 133w + 502)x^3 + (57w^2 + 379w + 56)x^2\\
& + (488w^2 + 94w + 185)x + 265w^2+ 385w + 137\bigr)\Bigr)/\mathbf{p}(x),
\end{align*}

\begin{align*}
\mathbf{C}_5(x,y)&=\Bigl(\bigl((112w^2 + 445w + 474)x^{10} + (180w^2 + 135w + 266)x^9 + (216w^2 + 277w + 100)x^8\\
& + (320w^2 + 277w + 218)x^7+ (329w^2 + 62w + 114)x^6 + (228w^2 + 71w + 411)x^5\\
& + (211w^2 + 9w + 353)x^4 + (45w^2 + 486w + 43)x^3 +(351w^2 + 464w + 377)x^2\\
& + (137w^2 + 340w + 409)x + 148w^2 + 487w + 454\bigr)+y\bigl((480w^2 + 393w + 429)x^9\\
& + (451w^2 + 335w + 482)x^8 + (14w^2 + 6w + 516)x^7 + (30w^2 + 462w + 430)x^6\\
& +(359w^2 + 353w + 248)x^5 + (193w^2 + 313w + 384)x^4 + (281w^2 + 510w + 111)x^3\\
& + (336w^2 + 124w + 159)x^2 +(492w^2 + 374w + 176)x + 184w^2 + 12w + 66\bigr)\Bigr)/\mathbf{p}(x),
\end{align*}

\begin{align*}
\mathbf{C}_6(x,y)&=\Bigl(\bigl((449w^2 + 288w + 365)x^{11} + (320w^2 + 375w + 357)x^{10} + (262w^2 + 70w + 425)x^9 \\
&+ (214w^2 + 312w + 445)x^8+ (302w^2 + 293w + 211)x^7 + (111w^2 + 24w + 333)x^6 \\
&+ (218w^2 + 328w + 127)x^5 + (w^2 + 252w + 419)x^4 +(286w^2 + 442w + 99)x^3 \\
&+ (217w^2 + 32w + 118)x^2 + (142w^2 + 61w + 406)x + 466w^2 + 251w + 287\bigr)\\
&+ y\bigl((283w^2 + 185w + 37)x^9 + (465w^2 + 226w + 378)x^8 + (210w^2 + 95w + 207)x^7 \\
&+ (432w^2 + 59w + 496)x^6 +(159w^2 + 479w + 450)x^5 + (201w^2 + 273w + 129)x^4\\
& + (494w^2 + 100w + 424)x^3 + (456w^2 + 349w + 214)x^2 +(476w^2 + 176w + 515)x \\
&+ 90w^2 + 343w + 520\bigr)\Bigr)/\mathbf{p}(x)
\end{align*}
with
\begin{align*}
\mathbf{p}(x)&=x^{10} + (433w^2 + 126w + 135)x^9 + (3w^2 + 477w + 283)x^8 + (476w^2 + 507w + 488)x^7 \\
&+(308w^2 + 158w + 387)x^6 + (104w^2 + 313w + 353)x^5 + (371w^2 + 193w + 173)x^4 \\
&+(452w^2 + 454w + 409)x^3 + (21w^2 + 196w + 16)x^2 + (164w^2 + 97w + 323)x \\
& + 503w^2 + 458w + 485.
\end{align*}

Now take $P=(7, 155w^2 + 306w + 310)$ and $Q=1987P=(311w^2 + 393w + 38, 418w^2 + 167w + 33 )$. Then
$S=\pi(P,P^\sigma,P^{\sigma^2})=\varphi(P)+\varphi(P)^\sigma+ \varphi(P)^{\sigma^2}=[D_S-3(\infty_1)]$
with $D_S$ cut by
\[x^3 + 364x^2z + 3xz^2 + 380z^3, y^3 + 340y^2z + 302yz^2 + 190z^3,\]
and $T=\pi(Q,Q^\sigma,Q^{\sigma^2})=\varphi(Q)+\varphi(Q)^\sigma+ \varphi(Q)^{\sigma^2}=[D_T-3(\infty_1)]$
with $D_T$ cut by
\[x^3 + 205x^2z + 158xz^2 + 25z^3,y^3 + 108y^2z + 424yz^2 + 491z^3.\]
We checked that $T=1987S$.

\section{Conclusion}
In this paper, we have given an algorithm for computing $(\ell,\ell,\ell)$-isogenies from the Weil restriction of an elliptic curve to the Jacobian of a non-hyperelliptic curve.
The time-consuming part of the algorithm comes from evaluating the Weil functions on the codomain of the isogeny, so the complexity (in terms of the prime $\ell$) of the algorithm is $\tilde{O}(\ell^3)$.
This algorithm can be used to transform the discrete logarithm problem in elliptic curves over cubic extension fields $\mathbb{F}_{q^3}$ into the corresponding problem in the Jacobian of a
non-hyperelliptic curve over $\mathbb{F}_q$, where the problem can be solved in a time of $\tilde{O}(q)$. This method gives a positive answer to the question whether elliptic curves over $\mathbb{F}_{q^3}$
with prime order can be attacked with cover attacks.


\bibliographystyle{abbrv}
\bibliography{cabibtex}

\end{document}